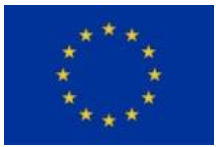
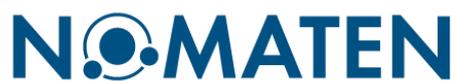

This work was carried out in whole or in part within the framework of the NOMATEN Centre of Excellence, supported from the European Union Horizon 2020 research and innovation program (Grant Agreement No. 857470) and from the European Regional Development Fund via the Foundation for Polish Science International Research Agenda PLUS program (Grant No. MAB PLUS/2018/8), and the Ministry of Science and Higher Education's initiative "Support for the Activities of Centers of Excellence Established in Poland under the Horizon 2020 Program" (agreement no. MEiN/2023/DIR/3795).

The version of record of this article, first published in Applied Spectroscopy, 2024, 0(0), is available online at Publisher's website:
https://dx.doi.org/10.1177/00037028241292087

# Using Label-Free Raman Spectroscopy Integrated with Microfluidic Chips to Probe Ferroptosis Networks in Cells

Muhammad Muhammad [a,b], Chang-Sheng Shao [a,c], Raziq Nawaz [a,d], Amil Aligayev [a,d,f], Muhammad Hassan [e], Mona Alrasheed Bashir [a,d], Jamshed Iqbal [a,g], Jie Zhan [b], Qing Huang [a,d]*

[a] CAS Key Laboratory of Ion-Beam Bioengineering, Institute of Intelligent Machines, Hefei Institutes of Physical Science, Chinese Academy of Sciences, Hefei 230031, China

[b] CAS Key Laboratory of Atmospheric Optics, Anhui Institute of Optics and Fine Mechanics, Chinese Academy of Sciences, Hefei 230031, China

[c] CAS High Magnetic Field Laboratory, Hefei Institutes of Physical Science, Chinese Academy of Sciences, Hefei 230031, China

[d] Science Island Branch of Graduate School, University of Science and Technology of China, Hefei 230026, China

[e] Intelligent Nanomedicine Institute, The First Affiliated Hospital of USTC, Division of Life Sciences and Medicine, University of Science and Technology of China, Hefei 230027, China

[f] NOMATEN Centre of Excellence, National Center for Nuclear Research, 05-400 Swierk/Otwock, Poland

[g] Center for Advanced Drug Research, COMSATS University Islamabad, Abbottabad Campus, Abbottabad 22060, Pakistan

* **Corresponding author:**

Prof. Dr. Qing Huang, huangq@ipp.ac.cn, ORCID: 0000-0002-8884-2064

## Abstract

Ferroptosis, a regulated form of cell death driven by oxidative stress and lipid peroxidation, has emerged as a pivotal research focus with implications across various cellular contexts. In this study, we employed a multifaceted approach, integrating label-free Raman spectroscopy and microfluidics to study the mechanisms underpinning ferroptosis. Our investigations included the ferroptosis initiation based on the changes in the lipid Raman band at 1436 cm$^{-1}$ under different cellular states, the generation of reactive oxygen species (ROS), lipid peroxidation, DNA damage/repair, and mitochondrial dysfunction. Importantly, our work highlighted the dynamic role of vital cellular components, such as NADPH, ferredoxin clusters, and key genes like GPX-4, VDAC2, and NRF2, as they collectively influenced cellular responses to redox imbalance and oxidative stress. Quantum mechanical (QM) and molecular docking simulations (MD) provided further evidence of interactions between the ferredoxin (containing 4Fe-4S clusters), NADPH and ROS which led to the production of reactive Fe species in the cells. As such, our approach offered a real-time, multidimensional perspective on ferroptosis, surpassing traditional biological methods, and providing valuable insights for therapeutic interventions in diverse biomedical contexts.

## Introduction

Ferroptosis, an iron-dependent form of regulated cell death, has garnered increasing attention due to its involvement in various pathophysiological processes including neurodegenerative disorders, cancer, and tissue damage caused by oxidative stress.[1–3] Understanding the metabolic pathways and molecular mechanisms underlying ferroptosis is pivotal for unraveling its roles in various contexts and holds potential for therapeutic interventions.[4,5] In cancer biology, a specific challenge pertains to comprehending the dynamics of ferroptosis and its implications for therapeutic intervention.[6] For example, despite their diverse functions, Fe–S clusters are most commonly recognized for their critical role in electron transfer pathways essential to processes like photosynthesis and mitochondrial respiration. In mitochondria, Fe–S centers are crucial for the tricarboxylic acid cycle (TCA) and the electron transport chain (ETC).[7] In the ETC, electrons donated by NADH and FADH2 pass through Fe–S clusters present in Complexes I, II, and III, ultimately reducing molecular oxygen to water. However, when cells experience high levels of ROS, excess mitochondrial Fe can catalyze further ROS production or disrupt enzyme activity. This dysregulation of mitochondrial iron metabolism has been linked to ferroptosis.[8] Furthermore, NADPH plays a critical role in mitigating oxidative stress by providing reducing power necessary for the regeneration of antioxidants, such as glutathione, which neutralize ROS. In a ROS-enriched environment, NADPH disruption elevates glutathione peroxidase, which detoxifies lipid hydroperoxides.[9] However, when NADPH levels are inadequate, the antioxidant defense is compromised, leading to increased lipid peroxidation. This process involves the oxidative degradation of lipids, resulting in the formation of harmful Fe (through Fenton reaction) that can further propagate oxidative damage and contribute to cell dysfunction or death.[10] Amidst the expanding recognition of ferroptosis's significance, the exact molecular determinants affecting the susceptibility or resistance of cancer cells to ferroptosis remain elusive. This ambiguity is particularly evident in the context of distinct cancer types and microenvironments, where the role of ferroptosis at different stages of cancer progression remains unclear. With this, a specific problem arises in identifying the molecular bridges that connect ferroptosis with pivotal cancer-associated pathways. This includes analyzing the oncogenic signaling, metabolic reprogramming, and redox imbalance causing ferroptosis.[11–13] A solution to this special problem could be molecular profiling that dissects these pathways and the ferroptotic process. Through systematic investigation of the genes, proteins, and metabolites that converge to ferroptosis, a comprehensive molecular blueprint could emerge. This could encompass predictive molecular signatures that guide therapeutic decisions for ferroptosis responsiveness of individual cancer patients.

In recent years, advancements in microfluidic technology and label-free vibrational spectroscopic tools have offered novel perspectives on studying cellular dynamics at unprecedented resolutions.[14–16] Over the years, by providing controlled release and enabling non-destructive molecular probing, these techniques have revolutionized the ability to investigate cellular responses to stimuli and perturbations.[16–21] In this study, we have employed a microfluidic chip coupled with label-free Raman spectroscopy to explore the dynamics of ferroptosis in cellular systems. Raman spectroscopy provided numerous advantages such as direct observation of change in cell state through variations in characteristic Raman bands of the cells enabling correlation of the changes in cellular environment with the Raman bands. Specifically, we have investigated the correlation between changes in lipid state in reactive oxygen species (ROS) environment, suppression of ROS through nicotinamide adenine dinucleotide phosphate (NADPH), and the impact of mitochondrial Fe-S protein clusters on redox imbalance. Our approach has elaborated a comprehensive time-dependent and concentration-dependent DNA damage and repair process in cells showcasing lipid peroxidation and redox imbalance. Our findings further revealed a connection between lipid peroxidation state causing the production of reactive $Fe^{2+}$ species and multiple cellular genes which regulated the dynamics of important biomolecules leading to ferroptosis. With the aid of QM and MD simulations, the interaction of ferredoxins with NADPH and ROS was studied, which elaborated our understanding of reactive Fe species generation in cells. The combination of microfluidics, label-free Raman microscopy and fluorescence microscopy empowered us to delve into the molecular interactions underlying ferroptosis.

## Materials and methods

### Y-channel microfluidic chip fabrication

The microfluidic chip was fabricated by WenHao Microfluidic Chip & Device Manufacturer (Suzhou, China). In brief, photolithography was employed to define the desired channel patterns on a photoresist-coated substrate. This was followed by transferring the pattern onto the resist. Next, a plasma etcher was utilized to transfer the pattern onto the PMMA substrate through dry etching. The microfluidic channels were created with a width of 100 μm and a depth of 50 μm. Subsequently, a bonding process was carried out, where a separate PMMA layer was aligned and bonded onto the patterned substrate. This was done using a vacuum sealing system or a thermal bonder. After bonding, a micro-milling machine was applied to create the access ports and outlets for fluid flow control. Finally, the assembled device underwent quality control measures to ensure the integrity of the channels and overall functionality.

**Cell flow through microfluidic channels**

The cell suspension was introduced into the channels using an automatic microfluidic syringe-injection system LST01-1A (HuiYu-WeiYe Microfluid Equipment, Beijing China) integrated with a precise microlitre fluid syringe (705 series, Hamilton (Shanghai) Laboratory Equipment Cp., Ltd). Following the reference protocols, the cells flowed through the transport pipes to the inlet 1 at the rate of 0.05 mL/min. Cell reagents including media, PBS, etc., were injected through inlet 2. The excess suspension in the microchannel was extracted from access port 3 using an outer connected microfluid syringe.

**Cell Culture and treatment**

Human lung carcinoma H1299 cells were maintained in Dulbecco's Modified Eagle Medium (DMEM) supplemented with 10% fetal bovine serum (FBS) and 1% penicillin-streptomycin at 37 °C in a humidified atmosphere containing 5% $CO_2$. For drug treatments, cells at 70-80% confluence were exposed to the ferroptosis-inducing drug or vehicle control (DMSO) for the indicated durations. For induced ferroptosis states, 25 mg/mL (5 µL) of NADPH (Beyotime Biotechnology Shanghai, China), 1 µg/mL (5 µL) of ferredoxin (4Fe-4S cluster protein, Sigma Aldrich Shanghai China) were added to the cells for 24 h prior to qPCR and Western blot. Regarding Raman experiments, 50 mg/mL of NADPH was added with cells (1:1 v/v) and spectral data were acquired.

**Raman spectra from cells and data treatment**

Raman measurements were performed using a confocal Raman spectrometer (XploRa Raman spectrophotometer Horiba Scientific, Japan) equipped with a 785 nm laser, a 10× objective, and a 600 lines/mm grating. The spectrometer was calibrated using the Raman signal of Si (520 $cm^{-1}$) and spectra were recorded by NG-LabSpec Raman software. For point spectra, the total integration time was 5 s with a laser power of 15 mW. The Raman imaging measurements were performed with an integration time of 10 s per point with a laser power of 57 mW, depending on the approximate size of the cell to complete one image. Spectral signals were denoised with Periodic Wavelet Transform algorithm under threshold level 1. The spectral data was tuned for analysis in commercial OriginPro-2019 (Academic) software (Origin Lab Corporation, Northampton, MA, USA). The spectral data were baseline corrected and smoothed using a 5th-degree polynomial and Savitzky-Golay smoothing filter. The Raman intensity was estimated by subtracting the absolute peak value from the average of the minima. The randomly acquired spectra were represented as mean±SD and an independent t-test was used for data comparison. Linear and non-linear fitting on data points was performed in built-in fitting algorithms in OriginPro-2019.

**Cell staining with ferroptosis probes**

Briefly, 24 h before treatment, ca. 6000 cells were plated in 96-well plates in a final volume of 100 mL, and then cultured for 24 h. Then cells were washed 3 times with PBS buffer before the addition of the fluorescence probe. And after adding the Liperfluor probe solution the plate was incubated for another 30 min. The nuclei were stained with Hoechst 33342 (Targetmol) and fluorescence density was measured using a microplate reader (SpectraMax M5, Molecular Devices, USA). All the images were obtained using 488 nm and 561 nm lasers to excite the ROS green probe and Liperfluo, while the 405 nm laser lines were used to excite Hoechst-33342. The High-Content Screening system Cell Insight CX5 HCS (Thermo Fisher, Waltham, MA) was used for automatic photo and quantitative analysis. The contrast and brightness of the image were adjusted by Cellomics software (Thermo Fisher, Waltham, MA).

**Western blot**

Cells were lysed in RIPA buffer (Biosharp, China) supplemented with protease and phosphatase inhibitors cocktail (Roche). Protein concentrations were determined using the BCA Protein Assay Kit (Pierce). Equal amounts of protein were separated by SDS-PAGE and transferred onto 0.45μm PVDF membranes (BioRad, USA). Membranes were blocked with 5% non-fat milk for 1 hour at room temperature and then incubated with primary antibodies overnight at 4°C. After washing, membranes were incubated with HRP-conjugated secondary antibodies for 1 hour at room temperature. Protein bands were visualized using the ECL Western Blotting Detection System (Oxford Bio, UK) and quantified using ImageJ software.

**RNA Extraction and Quantitative Real-time PCR (qPCR)**

Total RNA was isolated using the TRIzol reagent (Invitrogen) following the manufacturer's instructions. The concentration of RNA was determined using a NanoDrop spectrophotometer (ThermoFisher Scientific, USA). cDNA was synthesized from 1 μg of total RNA using the

SuperScript III Reverse Transcriptase kit (Invitrogen). Next, qPCR was performed using the SYBR Green PCR Master Mix (Yisheng, China) on a Roche 480 Real-Time PCR System (Roche, Swiss). The relative expression levels of target genes were normalized to GAPDH and calculated using the $2^{-\Delta\Delta Ct}$ method.

**Intracellular ROS production assay**

2',7'-dichlorofluorescein diacetate (DCFH-DA) was utilized to assess the intracellular generation of ROS. Briefly, H1299 cells were adherently grown in 6-well plates overnight and after replacing with fresh media containing ROS ($H_2O_2$: 1 mM, 5 µL/well), the cells were cultivated for another 24 h. After washing the cells with PBS three times, they were incubated with DCFH-DA (1 mM, 10 µL) for 30 min. The cells were repeatedly washed with Hanks Balanced Salt Solution (HBSS) buffer and dissolved in PBS. Finally, a fluorescence microscope (EVOS FL ThermoFisher Scientific, USA) was employed to assess the ROS content ($\lambda_{ex}$ = 488 nm, $\lambda_{em}$ = 525 nm for DCFH-DA).

**Quantum chemistry calculations**

The simulations for the present work were performed using Gaussian16.[22] NADPH along with all the possible tautomers was drawn using the GaussView 06 and optimized without any geometry constraint. To determine the relative stability of the structure, the optimizations of the tautomers were carried out with parameters of pure functional of Perdew, Burke, and Ernzerho PBE1PBE/aug-cc-pvdz.[23–25] The atoms in the tautomers were found in the 1st and 2nd rows of the periodic table. The aug-cc-pvdz algorithm containing 1s, 1p, and 1d functions was chosen for fine-tuned optimization of the tautomers. The PBE1PBE method included a PBE exchange correlation with a hybrid parameter. Regarding theoretical accuracy, the vibrational analysis was performed for the confirmation of true minima (no imaginary frequency) in the structure. After the structural optimization, the relative energies were computed to identify the most stable and optimal tautomer in the liquid phase.

**Molecular Docking**

Molecular docking simulations were performed to investigate the interactions between Ferredoxin (4Fe-4S), NADPH, and $H_2O_2$ molecules. The 3D structure of Ferredoxin was obtained from the Protein Data Bank (www.rcsb.org). The structure of NADPH was optimized using Gaussian16. The docking simulations were carried out using SwissDock (www.swissdock.ch/docking), an online platform provided by the Swiss Institute of

Bioinformatics, to explore the potential binding interactions between the optimized NADPH ligand and the Ferredoxin (4Fe-4S) receptor. The output files generated from these simulations were further analyzed using Chimera 1.17.3, enabling the visualization and examination of the ligand-receptor interactions. These interactions were illustrated through graphical images, providing valuable insights into the molecular basis of the binding between Ferredoxin (4Fe-4S) and NADPH.

## Results and discussion

### Control verification of redox imbalance and NADPH as ROS suppressor

The experimental design involving a microfluidic channel and Raman setup is demonstrated in Figure 1a. We performed control experiments in cells for the induction of ROS through controlled exposure to small volumes of hydrogen peroxide since ROS, as pivotal mediators of oxidative stress and signaling, play a crucial role in cellular homeostasis.[26,27] We observed a disparity between ROS levels in blank and ROS-induced cells (Figure 1b-1c) confirming the observable ROS within the cellular microenvironment, thus validating a robust foundation for subsequent investigations. We then inserted ROS-treated cells into micro channels through an auto-pressure syringe system and utilized label-free Raman spectroscopy to capture spectral changes in both blank and ROS-induced cells, which recorded varying molecular signatures specifically at 1375 and 1436 cm$^{-1}$, a region associated with lipid peroxidation, the spectrum provided insights into the direct consequences of ROS-mediated oxidative stress.[28] However, the incorporation of NADPH as a modulator to regulate ROS in cells provided a layer of depth to the provided NADPH's pivotal role in cellular redox balance and its association with the early-stage antioxidant defense system.[29,30] We found that through NADPH level variations directly influenced ROS levels in a concentration-dependent manner through changes in the lipid band at 1436 cm$^{-1}$ (Figure 1d). We noticed that decreasing NADPH concentrations escalated ROS levels, which downplayed the observed lipid band changes. One possible reason could be chemical interaction which allows to suppress the ROS levels within cells stems from its involvement in the regeneration of reduced glutathione (GSH), a powerful antioxidant.[31–33] NADPH fuels the glutathione reductase enzyme, which catalyzes the reduction of oxidized glutathione (GSSG) back to its reduced form (GSH). This enzymatic reaction, facilitated by NADPH, directly counteracts the accumulation of ROS by restoring the antioxidative capacity of GSH.[34] The ramifications encompass a mechanistic elucidation of the relationship between ROS and lipid alterations, thereby augmenting the correlation between Raman observations and cellular responses. Furthermore, we performed fluorescence imaging to validate the impact of NADPH levels on ROS (Figure 1e). This fluorescence-based evidence reinforced the mechanistic understanding established through Raman spectroscopy, collectively underscoring the integral role of NADPH in regulating ROS dynamics within the cellular microenvironment. However, concerning the controlled alterations in ROS, we hypothesized that the observed continuous modulation of ROS levels (Figure 1f), may couple with the redox imbalance within the cells, aligning with the time-dependent exposure to NADPH and subsequent accumulation of NADPH molecules within the cells.

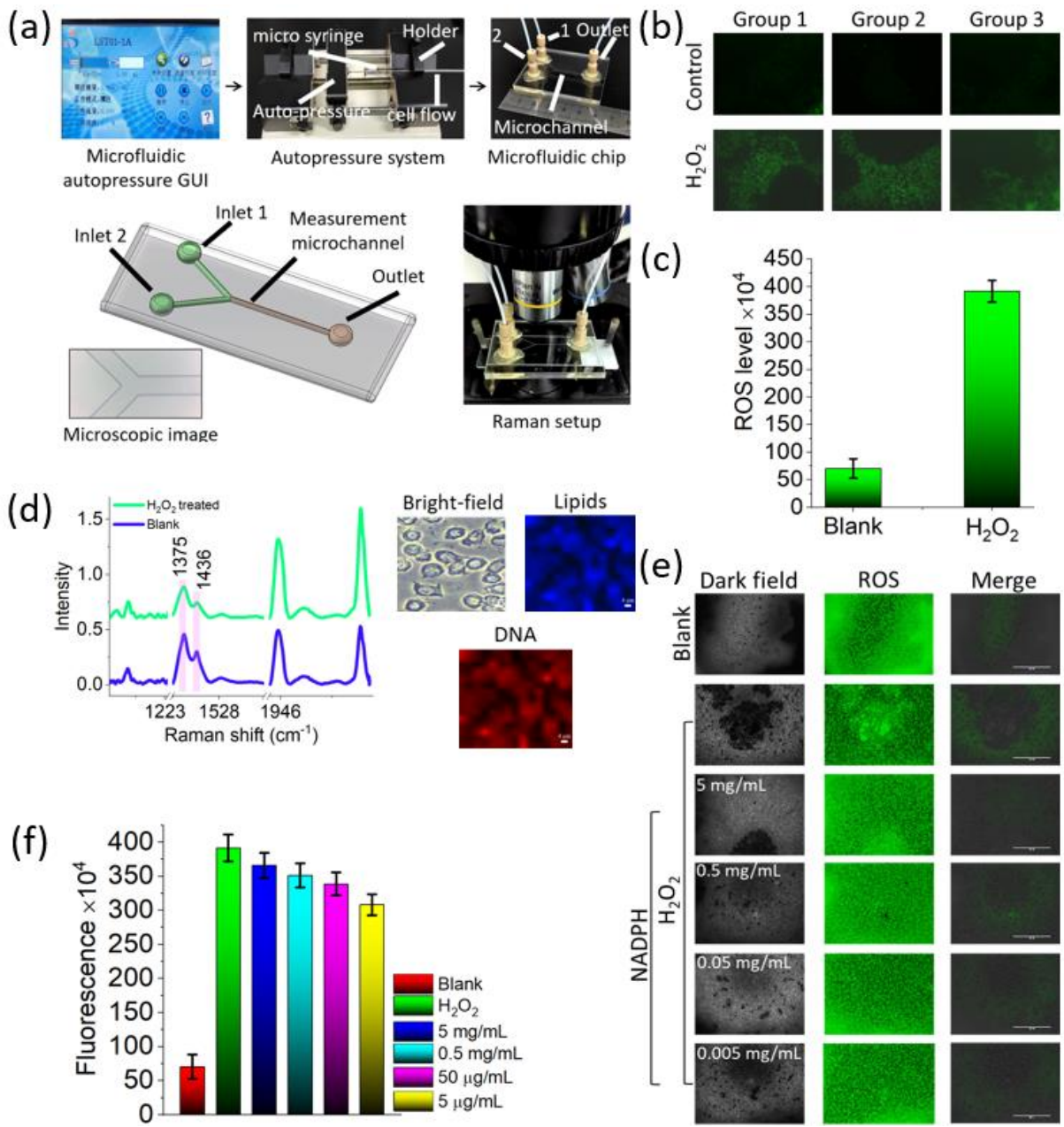

Figure 1: (a) Microfluidic chip integrated Raman measurement setup. (b) Fluorescence microscopy images depicting the augmentation of ROS levels within H1299 cells following

induction. (c) Quantification of ROS expression levels derived from the analysis of cells as presented in (b). (d) Raman spectra captured from the cells after ROS induction, elucidating molecular alterations. (e-f) Fluorescence images and correlated intensity levels within the cells subjected to ROS induction, followed by controlled NADPH release treatment.

**Raman signatures of NADPH-mediated dynamic cellular responses**

The analysis of Raman spectra acquired at different time intervals ranging from 0 h to 2 h-20 min provided time-dependent patterns of NADPH reaction with lipid and DNA (Figure 2a). We broke down the spectral information into two sections (Figure 2b) and individually analyzed the alterations in Raman bands at 1090, 1375, and 1436 $cm^{-1}$ which are related to lipid and cellular DNA.[35,36] Particularly, the observed intensity changes in the lipid band at 1090 $cm^{-1}$ (Figure 2c), (correlated with C-C stretching in lipids and O=P=O stretching in DNA), illustrated the interactions occurring between NADPH and lipids (supporting information Figure S1). The initially ascending trend of this band's intensity over the initial hour of NADPH interaction, followed by a sustained pattern at 140 min, suggested an early response possibly linked to cell signaling or adaptive adjustments to the redox environment.

*DNA damage and repair during redox imbalance:* The variations at 1375 $cm^{-1}$ (attributed to DNA bases) introduced a temporal dimension to our findings (Figure 2d). The initial dip in intensity within the first 80 min of NADPH interaction was indicative of DNA damage recognition and initiation of repair pathways. DNA repair processes, such as nucleotide excision repair (NER) or base excision repair (BER), often involve the temporary unwinding of DNA strands and excision of damaged bases.[37,38] This led to a decrease in the observed Raman intensity as the DNA structure changed to facilitate the repair process. Subsequently, the rising intensity of the 1375 $cm^{-1}$ band beyond the initial 80 min signified the restoration phase of the repair process. As damaged DNA segments were excised and replaced, the DNA strand underwent reconstitution, potentially leading to a recovery in the observed Raman intensity. This pattern aligns with the temporal progression of DNA repair pathways as suggested by Vikas and co-workers, where the role of NADPH oxidases was shown in DNA damage at short time points,[39] however, the continuous quantification of the DNA damage and following repair process was not achieved.

*Quantification of DNA repair during redox imbalance:* Considering the quantification of dynamic repair processes, the implications of the linear relationship between time and the Raman intensity of the 1375 $cm^{-1}$ band were noteworthy (Figure 2e). The linearity suggested a systematic alteration in the DNA structure, potentially mirroring the stepwise progression of DNA repair. This consistency also signified the precision of the cellular response, implying a regulated process that unfolded over time. This behavior added depth to NADPH's influence on DNA-related processes and underscored the multifaceted nature of cellular adjustments. Another observation emerged from the lipid band-cytochrome ratio ($I_{1436}/I_{568}$), and fluctuations in this band offered a compelling narrative of cellular responses to NADPH interaction (Figure 2f). The initial increase within the first 90 min signified an adaptive lipid remodeling in response to NADPH-driven redox changes. The subsequent drop at the 100$^{th}$ minute introduced an inflection point, possibly indicating a threshold beyond which the cellular machinery experienced a transition or modulation. The subsequent gradual and linear increase between the 100$^{th}$ and 130$^{th}$ min suggested a fine-tuning mechanism, wherein the cellular environment may have refined its lipid composition in response to prolonged NADPH interaction,[40] which was another confirmation of DNA process quantification obtained through Figure 2e. The eventual stabilization between the 130$^{th}$ and 140$^{th}$ min reflected an equilibrium attained in the cellular response which was indicative of a threshold reached in the cellular adaptation process.

The NADPH concentration-dependent exposure in ROS environment, elaborated a complex interplay between cell machinery, DNA, and lipid dynamics (Figure 2g). The initial decrease in intensity of this band in the ROS positive control suggested an early impact of oxidative stress on lipid and potentially DNA components. We no NADPH levels enable antioxidant systems, such as the GPX and thioredoxin systems, to prevent further oxidative damage.[41] The further decrease in intensity with lower concentrations of NADPH (5 µg/mL and 50 µg/mL) indicated insufficient NADPH to counteract ROS-induced damage. NADPH supports antioxidant defense systems by reducing glutathione (GSH), which neutralizes ROS.[42] Low NADPH levels result in insufficient GSH reduction, weakening the cellular defense against oxidative stress. However, the increase in intensity as NADPH concentration escalates beyond these levels hints at the NADPH in regulating cellular responses and maintaining DNA integrity (Figure 2h).

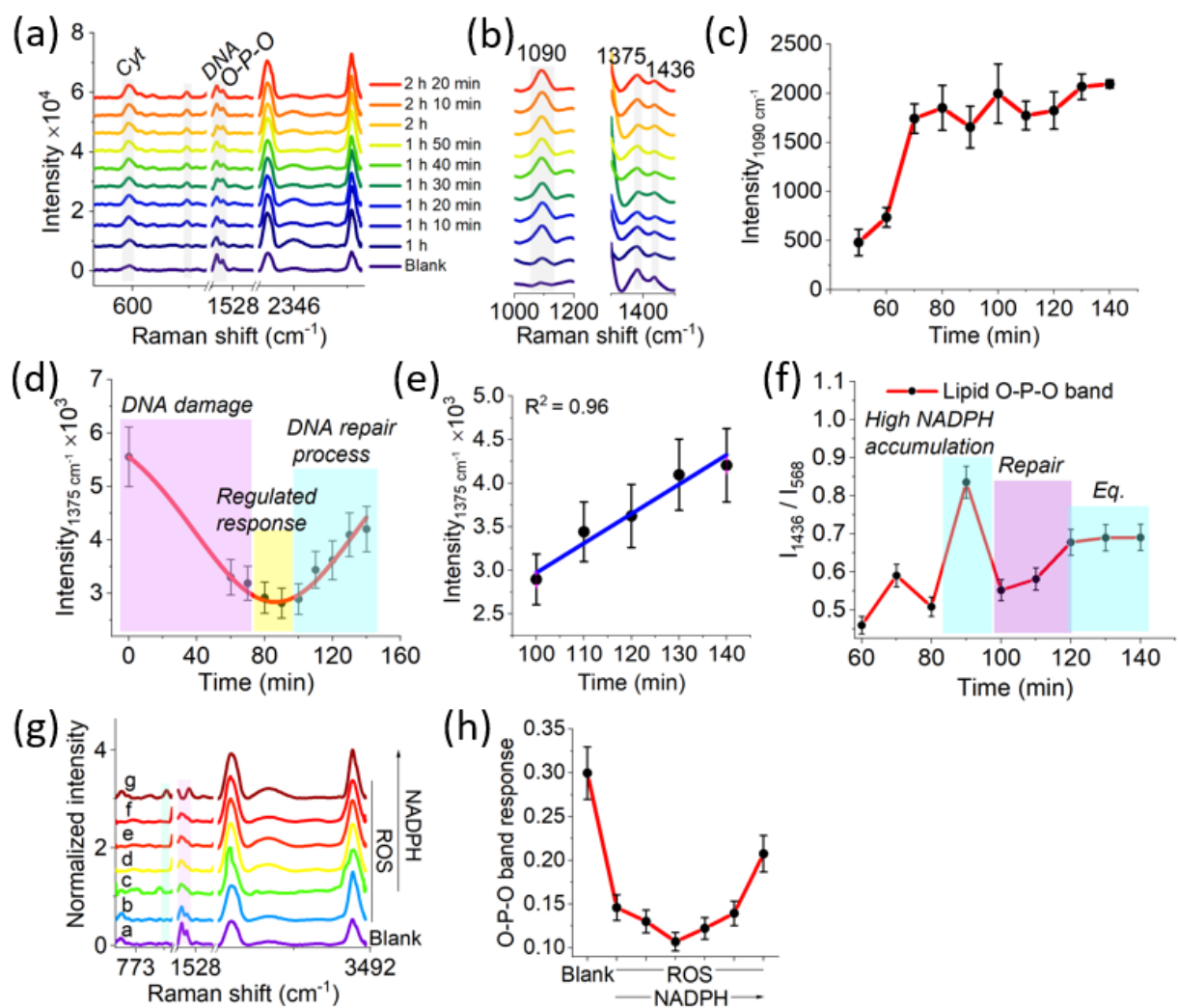

Figure 2: NADPH-driven dynamics, regulation of redox equilibrium and lipid Raman signatures. (a) Temporal Raman spectra portraying the evolving lipid and DNA Raman signatures during sequential exposure to NADPH. (b) Notable fluctuations in intensity within DNA and lipid bands, elucidating the intricate interplay instigated by NADPH interactions. (c) Time-dependent correlation between Raman intensity at 1090 cm$^{-1}$ (C-C stretching in lipids and O=P=O stretching in DNA) and the progression of exposure time. (d) Temporal association between Raman intensity at 1375 cm$^{-1}$ (DNA-associated) and the temporal course of exposure. (e) Linear relationship between $I_{1375 \text{ cm}^{-1}}$ and time. (f) Chronological correspondence between Raman intensity ratio ($I_{1436}/I_{568}$), reveals distinctive junctures of NADPH accumulation, DNA repair processes, and eventual equilibrium dynamics. (g) Raman spectra capturing the responses elicited post-ROS exposure as a positive control, along with concentration-dependent responses following interaction with NADPH (a-f: Blank, ROS control, c-g: NADPH 5 µg/mL – 50 mg/mL). (h) Concentration-dependent responses of the O-P-O band in cells after NADPH treatment.

**Correlation of Lipid Raman band with Fe-S clusters driven cellular responses**

Initiated with ROS induction, $H_2O_2$-treated cells exhibited elevated ROS expression through fluorescence microscopy (Figure 3a-3b). Subsequently, Fe-S-treated cells showed a contrasting trend, showing a decline in ROS expression relative to the blank. Intriguingly, the co-treatment of cells with $H_2O_2$ and Fe-S sustained comparable ROS levels, yet significantly lower than the blank. From NADPH and Fe-S interaction, ROS expression showed a moderate increment compared to the prior scenario. The integration of NADPH with both $H_2O_2$ and Fe-S yielded ROS levels akin to the former instance, yet strikingly lower than the blank condition.

The Raman spectral analysis was performed concerning the lipid Raman band at 1436 $cm^{-1}$ (Figure 3c). The lipid band intensity, relatively stable in ROS-containing cells, showed a decrease upon Fe-S co-treatment. One possible reason is that Fe-S clusters, particularly in their damaged or altered states, could potentially interact with lipids in a way that may affect their structural integrity or composition.[43,44] This interaction might have led to lipid modifications or perturbations that manifested as changes in the lipid Raman band's intensity. The decrease in lipid band intensity reflected alterations in the lipid structure, potentially related to oxidative modifications or other Fe-S-induced changes. Furthermore, the incorporation of NADPH with Fe-S induced a further dip in lipid band intensity, albeit persisting at a notably lower level than the baseline blank (Figure 3d). As shown in Figure 3e, it showed a consistent inverse correlation between ROS expression and lipid band intensity across diverse scenarios. As ROS expression escalated, the intensity of the lipid band underwent a discernible decrease.

*Fe-S cluster damage and $Fe^{2+}$ species:* The observed decline in ROS expression subsequent to Fe-S treatment was emblematic of the intricate interplay between Fe-S clusters and redox equilibrium. We doubted that the damage to Fe-S clusters might have disrupted their role in cellular redox homeostasis, prompting the release of Fe ions. These Fe ions, upon encountering $H_2O_2$ within the cellular environment, might have catalyzed the generation of reactive Fe species through Fenton reactions,[45] thereby amplifying oxidative stress and contributing to redox imbalance. We proved this hypothesis by first performing a Western blot analysis of GPX4 and xCT genes (which were activated to counter the induction of reactive Fe ions) and observed the changes in protein levels (Figure 3f).

*Activation of FTH-1 and GPX-4 genes:* In gene expression analysis, we examined the FTH-1 gene across various cell treatment scenarios associated with ferroptosis (Figure 3g). We found distinct patterns of FTH-1 expression in response to different treatments. Specifically, the combination of NADPH and $H_2O_2$ resulted in a slight increase in FTH-1 expression compared to blank cells, indicating a countering of oxidative stress. Furthermore, NADPH combined with ferredoxin clusters led to even higher FTH-1 levels, suggesting an enhanced cellular response to Fe ions related stress. The introduction of ROS containing cells alongside NADPH and ferredoxin initially reduced FTH-1 expression but was followed by a substantial increase in FTH-1 levels in $H_2O_2$ and ferredoxin-treated cells. This modulation of FTH-1 expression underscored its potential involvement in ferroptosis regulation, where cellular responses to oxidative stress and iron homeostasis appear to be finely balanced, with implications for cell fate and survival. Similarly, we also investigated the GPX-4 gene expression. GPX-4 is a key player in protecting cells against lipid peroxidation and ferroptosis, making it a pivotal component in cellular defense mechanisms. Most importantly, the highest GPX-4 expression levels were recorded in $H_2O_2$ + ferredoxin-treated cells.

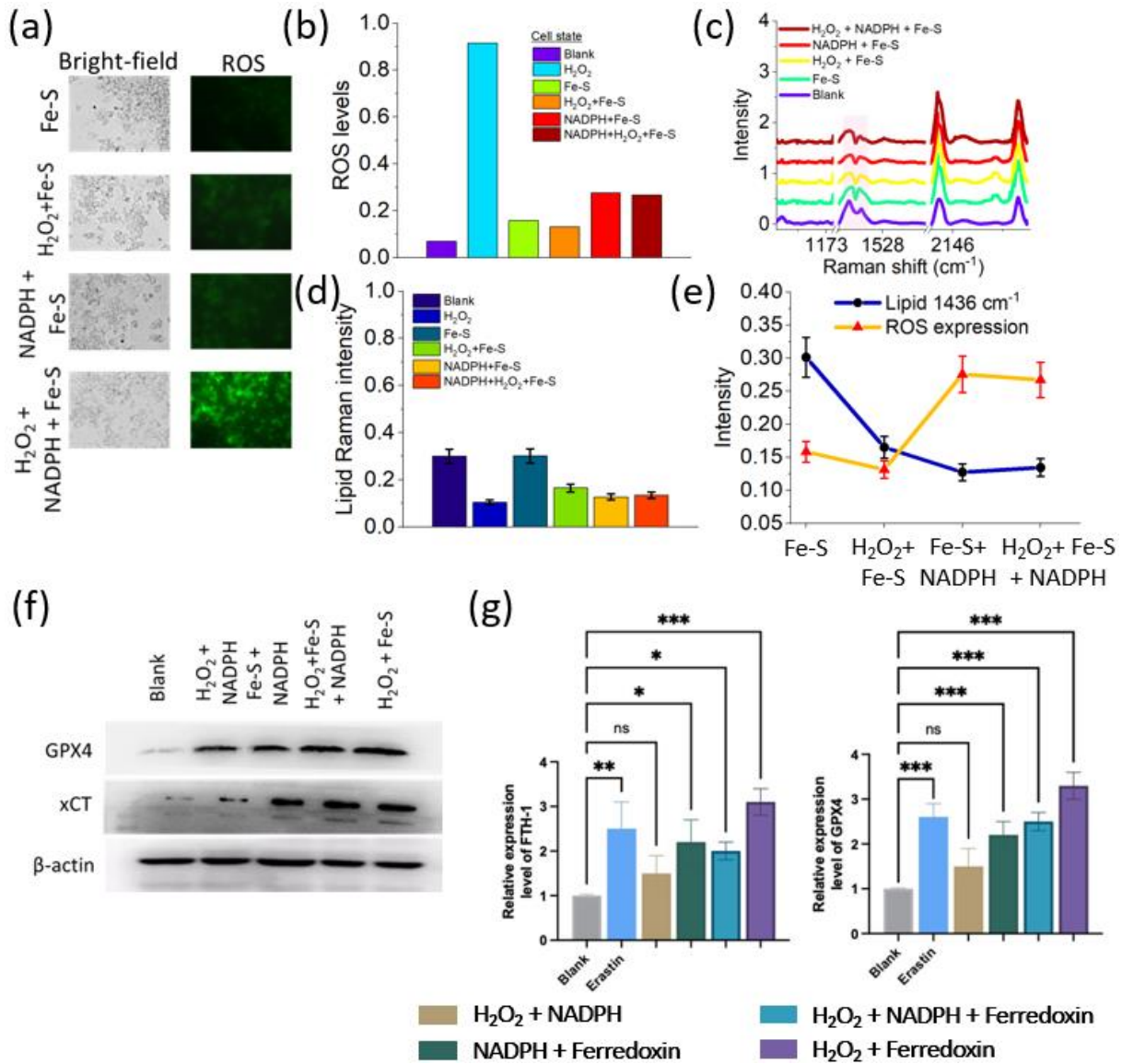

Figure 3: (a-b) Fluorescence microscopy images and ROS expression within distinct conditions. (c) Raman spectra obtained from cells under different conditions, elucidating intricate shifts in molecular profiles. (d) Raman spectral analysis of the lipid band at 1436 cm$^{-1}$ across diverse cellular contexts. (e) An inverse correlation between ROS expression levels and lipid band intensity across distinct cellular scenarios. (f) Western blot analysis of protein expression changes in cellular Fe-stressed environment. (g) FTH-1 and GPX-4 gene expressions in different cell states. P-values for significance levels: 0.12 (ns), 0.033 (*), 0.002 (**), <0.001 (***).

**Fe-S cluster interactions produce Fe$^{2+}$ and lead to ferroptosis**

The generation of reactive Fe species was validated into five cell groups including blank cells, ROS-induced cells, H$_2$O$_2$+NADPH-treated cells, H$_2$O$_2$+NADPH+Fe-S-treated cells, and H$_2$O$_2$+Fe-S reference. The fluorescence microscopy images from these groups provided a comprehensive view of Fe-S cluster involvement and its aftermaths (Figure 4a). Notably, the combined dosage of H$_2$O$_2$, NADPH, and ferredoxin-treated cells showed the highest levels of Fe ions. In the presence of ROS, ferredoxin clusters were vulnerable to oxidative damage. The subsequent inclusion of NADPH provided a regenerative capacity of ferredoxin to an extent. However, the presence of Fe-S clusters in this redox-imbalanced environment led to the release of Fe ions due to partial cluster damage. This released Fe contributed to the generation of reactive Fe species, resulting in elevated Fe$^{2+}$ levels.

To further validate this, the standard quantification of lipid peroxidation and reactive Fe species was performed (Figure 4b-4c). The elevation of Fe ions observed in the H$_2$O$_2$+NADPH+Fe-S-treated cells aligned with the increased levels of reactive Fe species in fluorescence images, confirming the proposed interaction between ferredoxin, ROS, and NADPH. Interestingly, quantum mechanical calculation showed 8 different conformers of NADPH with different stabilization energies (supporting information Figure S2). The stabilization energy of conformers was estimated and the most stable structure of NADPH (Figure 4d) was chosen for docking with Fe-S clusters (supporting information Figure S3). Similarly, Fe-S was optimized and its interaction with ROS and NADPH was separately analyzed with docking. As obvious in Figure 4e, the ROS and NADPH potentially interact with Fe binding site of Fe-S, leading to the change of state of Fe molecules. For detailed analysis, the site-by-site interaction of NADPH with amino acids of Fe-S was verified (supporting information Figure S4) binding site viable to charged ions.

***Regulation of VDAC2 and NRF2 genes:*** VDAC2 played a significant role in mitochondrial function and which was integral to regulating mitochondrial voltage and Fe ions transport. We found that the elevation of VDAC2 gene expression in response to NADPH+$H_2O_2$ treatment responded to counteract oxidative stress-induced mitochondrial perturbations (Figure 4f). The similarity in VDAC2 expression levels between this treatment and Erastin (as positive control) suggested a potential link between ferroptosis and alterations in mitochondrial function. Subsequent treatments with NADPH+Fe-S and NADPH+$H_2O_2$+Fe-S did not result in significant changes in VDAC2 expression, while $H_2O_2$+Fe-S treatment led to a slight decrease in VDAC2 levels. This reduction signified an adaptive response to restore mitochondrial homeostasis or indicate complex regulatory mechanisms. In the case of NRF2, the NADPH+$H_2O_2$ treatment led to a modest increase in NRF2 gene expression, confirming an initial response to oxidative stress. Notably, NADPH+Fe-S treatment induced even higher NRF2 expression, surpassing the levels observed in the blank cells and NADPH + $H_2O_2$-treated cells. This suggested that the presence of Fe-S clusters may amplify the cellular antioxidant response, potentially to counteract oxidative damage. Subsequently, the NADPH+$H_2O_2$+Fe-S treatment resulted in a further increase in NRF2 expression, signifying a synergistic effect of NADPH, ROS induction, and Fe-S clusters on NRF2 activation. The most substantial NRF2 expression levels were observed in cells treated with $H_2O_2$ + Fe-S, highlighting the potential significance of reactive Fe species produced during ferredoxin state variation in enhancing NRF2-mediated antioxidant responses.

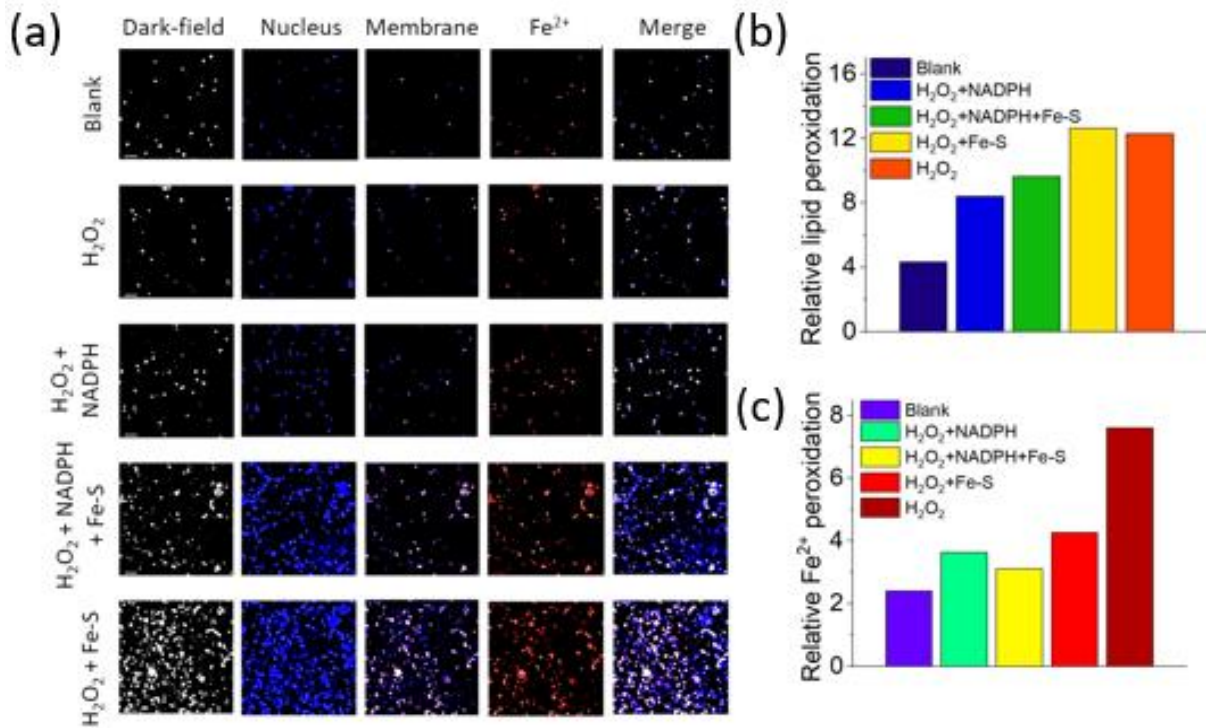
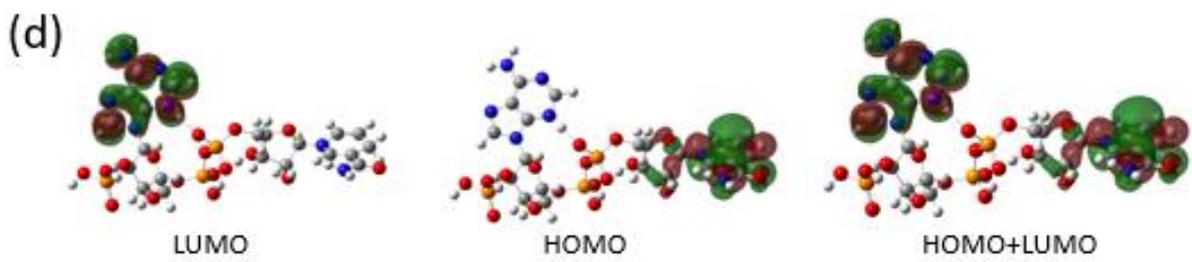
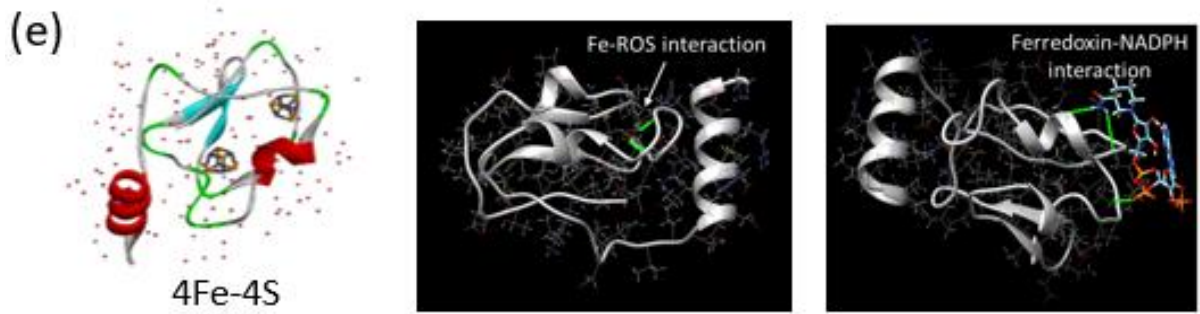
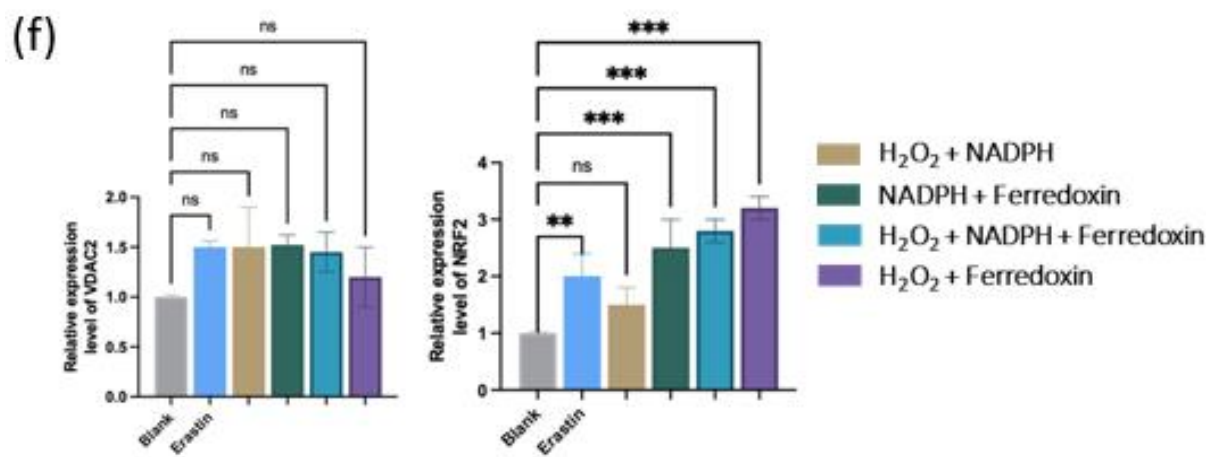

Figure 4: Cellular redox dynamics with Fe-S clusters damage and production of reactive Fe species. (a) Fluorescence imaging and microscopy encompass diverse cellular scenarios: blank cells, ROS control group, $H_2O_2$+NADPH-treated cells, $H_2O_2$+NADPH+Fe-S-treated cells, and $H_2O_2$+Fe-S reference group. (b) Quantification of lipid peroxidation through scenarios stated in (a) and estimation of peroxidation levels in lipids using fluorescence microscopy. (c) Quantification of reactive $Fe^{2+}$ species levels in different cell states, unveiling the intricate redox and Fe-S dynamics at play within the cellular environment. (d) Optimized structure of NADPH molecule and Fe-S cluster in aqueous environment. (e) Geometrical models of protein-ligand interaction between ferredoxin and reactive $H_2O_2$, and NADPH conformer. (f) Expression levels of VDAC2 and NRF2 genes were obtained through qPCR corresponding to different cell states. P-values for significance levels: 0.12 (ns), 0.033 (*), 0.002 (**), <0.001 (***).

The study of DNA damage and repair processes using label-free Raman-microfluidics holds significant implications for understanding how cells respond to external stressors for example, drugs and radiation. DNA damage can arise from various sources, including ionizing radiation and chemotherapeutic agents, leading to genome mutations, cancer, or cell death if repair mechanisms fail. One of the examples is the DNA damage induced by bleomycin anticancer drug which makes transitional changes from B-DNA to A-DNA upon interaction with cells.[46] Another example is the double-strand break and repair process study in yeast cells using microfluidics technique.[47]

This label-free technique is a powerful tool as it allows real-time, direct detection of biochemical changes within DNA and lipids without the need for fluorescent tags or markers, providing a more natural insight into cellular processes. In this work, we have demonstrated the ability of this technique to capture not only the initial DNA damage but also the cell's intrinsic repair processes. Specifically, we showed that the activation of repair mechanisms, such as the role of NADPH in counteracting ROS-induced oxidative stress, helps neutralize free radicals and prevents Fe-based apoptosis pathways. Our findings reveal a time-dependent and concentration-dependent role of NADPH in mitigating DNA damage, which further highlights the complex interplay of cellular repair mechanisms. The insights gained from this work could provide a novel platform for testing therapeutic strategies aimed at enhancing DNA repair processes, thereby improving treatments for conditions such as cancer, where DNA damage plays a pivotal role.

## Conclusion

This work has demonstrated the molecular mechanisms of ferroptosis in connection with the change in NADPH enzyme level. By integrating rapid investigation through Raman spectroscopy-microfluidic platform, we provided real-time dynamic cellular responses to oxidative stress and redox imbalance through the generation ROS resulting in lipid peroxidation, DNA damage, and mitochondrial dysfunction. It further highlighted the roles of key cellular components, such as NADPH, ferredoxin (4Fe-4S) clusters, and critical genes including GPX-4, VDAC2, and NRF2, in modulating responses to redox imbalance influencing the ferroptosis. By bridging between advanced analytical techniques and cellular biology, our study not only contributes to our understanding of ferroptosis but also opens avenues for targeted interventions in diseases where this form of cell death is implicated.

While this work has provided valuable insights into the mechanisms of ferroptosis, some areas require further exploration. For example, a more extensive investigation into different pathways is necessary for identifying potential therapeutic targets. Moreover, exploring the effects of potential ferroptosis modulators and therapeutic strategies in preclinical models and clinical trials will further bridge the gap between basic research and clinical applications. Addressing these areas of future work will enhance our grasp of ferroptosis and its role in various diseases, ultimately advancing its potential for clinical interventions.

## CRediT authorship contribution statement

M.M. conceived the study, performed the experiments and formal analysis, and wrote the original manuscript. C.S.S. assisted in fluorescence microscopy on cells. R.N. and A.A jointly performed QM simulations. M.H. performed molecular dynamics simulations. M.A.B assisted in cell culturing. All the authors, including J.I. and J.Z. participated manuscript editing and discussion. contributed to extensive discussions. Q.H. supervised the work, conceived the study, and reviewed and edited the paper. All co-authors have read and approved the final manuscript for publication.

## Declaration of competing interests

The authors declare no competing interests.

## Acknowledgments

This work was supported by Research Fund for International Young Scientists (RFIS-1, E22ADI35501), the Postdoctoral International Exchange Program (E32AAD18), National